\documentclass[referee,useAMS,usenatbib]{mn2e}

\usepackage{amssymb}
\usepackage{epsf}
\usepackage{bm}

\newcommand{\MNRAS}{MNRAS}
\newcommand{\ApJ}{ApJ}
\newcommand{\rhomax}{\rho^{\rm max}}
\title[Seismic signatures of strange stars with crust]
    {
Seismic signatures of strange stars with crust
    }
\author[A. I. Chugunov]{A. I. Chugunov\thanks{E-mail: andr.astro@mail.ioffe.ru}
\\ Ioffe Physico-Technical Institute,
Politekhnicheskaya 26, 194021 Saint-Petersburg, Russia}

\begin{document}

\pagerange{\pageref{firstpage}--\pageref{lastpage}} \pubyear{2006}

\maketitle

\label{firstpage}

\begin{abstract}
We study acoustic oscillations (eigenfrequencies, velocity
distributions, damping times) of normal crusts of strange stars.
These oscillations are very specific because of huge density jump at
the interface between the normal crust and the strange matter core.
The oscillation problem is shown to be self-similar. For a low (but
non-zero) multipolarity $l$ the fundamental mode (without radial
nodes) has a frequency $\sim 300$ Hz and mostly horizontal
oscillation velocity; other pressure modes have frequencies $\gtrsim
20$~kHz and almost radial oscillation velocities. The latter modes
are similar to radial oscillations (have approximately the same
frequencies and radial velocity profiles). The oscillation spectrum
of strange stars with crust differs from the spectrum of neutron
stars. If detected, acoustic oscillations would allow one to
discriminate between strange stars with crust and neutron stars and
constrain the mass and radius of the star.
\end{abstract}

\begin{keywords}
 stars: neutron -- stars: oscillations.
\end{keywords}

\section{Introduction}\label{SecIntrod}

Strange stars are hypothetical compact stars which are built
entirely or almost entirely of strange quark matter (containing
light $u$, $d$, and $s$ quarks and possibly electrons). The
hypothesis of strange stars is based on the idea of \citet{Witten}
that the strange quark matter is the absolutely stable form of
matter even at zero pressure. The hypothesis cannot be definitely
confirmed or refuted by available theories and experimental data.
Strange stars are attracting permanent attention \citep[see, e.g., a
recent review of][]{Weber2005}. One cannot exclude that at least
some stars, which are currently thought to be neutron stars, are in
fact strange stars. The theories predict the existence of bare
strange stars (with the strange matter extended to the very surface)
and strange stars with a thin crust of normal matter extended not
deeper than the neutron drip point \citep*[e.g.,][]{Alcock1986}. The
normal crust may occur, for instance, due to accretion of normal
matter onto a bare strange star.

Strange stars of masses $M$ much lower than the ``canonical''
neutron-star mass, $M \sim 1.4\, M_\odot$, are predicted to have
radii $R \ll 10$ km, essentially smaller than the radii of neutron
stars. However, strange stars with $M \sim 1.4\, M_\odot$ have
nearly the same radii, $R \sim 10$ km, as typical neutron stars.

We will show that strange stars with crusts could be potentionally
distinguished from neutron stars by their oscillation spectra.
Oscillations of strange stars have been analyzed by a number of
authors. \citet*{Yip1999} studied three types of quadrupole
oscillations of strange stars and hybrid neutron stars. They found
some differences in oscillation frequencies of such stars. Also,
they demonstrated that the damping times of oscillations are
sensitive to the model of strange matter. Recently \citet{Benhar}
showed that the combined knowledge of the frequency of emitted
gravitational waves and the mass or radius of a compact object would
allow one to discriminate between a strange star and a neutron star
and set stringent bounds on the parameters of quark matter. Both
groups of authors analyzed global oscillations of strange stars.

In this paper we focus on pressure oscillations (p and f modes) of
strange star crusts. In our previous papers \citep{Chug2005,Chug06}
we have studied oscillations localized in a neutron star crust due
to large multipolarity $l \gtrsim 100$. Here we show that acoustic
oscillations in a strange star crust are very specific even for low
$l$ because of a huge density jump at the interface between the
normal crust and the quark core.

\section{Formalism}
\label{SecFormal}

Because a strange star crust is very thin
(a few hundred meters for a strange star
of mass $M \sim 1.4\,M_\odot$), the approximation
of plane-parallel layer can be
used. Then the space-time metric in the crust can be written as
\begin{equation}
\label{metric}
 {\rm d} s^2=c^2\,{\rm d} t^2-\,{\rm d} z^2 -R^2\,
 ({\rm d}\vartheta^2+\sin^2 \vartheta\,{\rm d}\varphi^2),
\end{equation}
where the local time $t$ and the local depth $z$ are related to the
Schwarzschild time $\widetilde{t}$ and the circumferential radius
$r$ by
\begin{equation}
\label{LocVar}
    t= \tilde{t} \, \sqrt{1-R_{\rm G}/R},\quad
    z=(R-r)/\sqrt{1-R_{\rm G}/R},
\end{equation}
$r=R$ is the circumferential radius of the stellar surface,
$\vartheta$ and $\varphi$ are spherical angles, $R_{\rm G}=2GM/c^2$
is the gravitation radius, and $M$ is the gravitational mass of the
star. The metric (\ref{metric}) is locally flat and allows
us to use the Newtonian hydrodynamic equations for a thin envelope
with the gravitational acceleration
\begin{equation}
    g=\frac{GM}{R^2\sqrt{1-R_{\rm G}/R}}.
\end{equation}
The pressure in the strange star crust is mostly determined by
degenerate electrons and is almost independent of temperature $T$.
Accordingly, we can use the same zero-temperature equation of state
(EOS) for the equilibrium structure of the crust and for
perturbations. We will neglect the buoyancy force and study pressure
modes. We will also neglect elastic stresses which are unimportant
for crust oscillations in the frequency range of interest (they
would be important for lower frequencies $\sim 1$~Hz). The
linearized hydrodynamic equations (for a non-rotating star) for the
velocity potential $\phi$ (so that the oscillation velocity is ${\bm
V}=-\nabla \phi$) can be rewritten as \citep[see, e.g., the
monograph by][]{LambBook}
\begin{equation}
\label{phi}
    \frac{\partial^2 \phi}{\partial t^2}=c_{\rm s}^2\Delta \phi
    +{\bm g}\cdot\nabla \phi,
\end{equation}
where $c_{\rm s}^2\equiv {\partial P_0}/{\partial \rho_0}$ is the
squared sound speed. The potential $\phi$ can be presented in the
form
\begin{equation}
    \phi=e^{\imath\omega t}\,Y_{lm}(\vartheta,\varphi)\,F(z),
\end{equation}
where $\omega$ is an oscillation frequency, and
$Y_{lm}(\vartheta,\varphi)$ is a spherical function (see, e.g.,
\citealt*{Varshalovich}). An unknown function $F(z)$ obeys the
equation
\begin{equation}
\label{F}
    {\frac{{\rm d} {^2F}}{{\rm d} {z^2}}}+\frac{g}{c_{\rm s}^2} {\frac{{\rm d} {F}}{{\rm d} {z}}}
    +\left(\frac{\omega^2}{c_{\rm s}^2}
    -k^2\right)F=0,
\end{equation}
where $k^2=l(l+1)/R^2$. The requirement of vanishing Lagrange
variation of the pressure at the surface implies $F(z)$ to be
bounded as $z\rightarrow 0$. Because the strange quark matter at the
interface between and crust and the quark core is very dense and
almost incompressible, we impose the condition $F^\prime(h)=0$,
which means the vanishing radial displacement at the crust bottom
$z=h$. In other words, we study oscillations of the crust alone.

Following standard prescription we will call the modes without
radial nodes of $F(z)$ by f modes, and the modes with nodes by p
modes.

In analogy with the oscillation problem for neutron star crusts
\citep{Chug06}, the
oscillation problem for strange star crusts
is self-similar.
Taking the equilibrium pressure $P$ as an independent variable in Eq.\
(\ref{F}) we come to the equation
for an eigenvalue $\lambda =\omega^2/g^2$; this equation
contains the scaling parameter $\zeta=k/g$. Accordingly, the
eigenfrequencies can be written as
\begin{equation}
    \label{scaling}
     \omega_k^2=g^2\,f_k(\zeta).
\end{equation}
Here, $f_k(\zeta)$ are functions which can be calculated
numerically. Note, that for $l=0$ (radial oscillations) the
eigenfrequency is proportional to the gravity $g$.

For the polytropic EOS, the squared sound speed is $c_s^2=gz/n$ and
Eq.\ (\ref{F}) transforms to
\begin{equation}
\label{F_polytrop}
    z\,{\frac{{\rm d} {^2F}}{{\rm d} {z^2}}}+n{\frac{{\rm d} {F}}{{\rm d} {z}}}
    +\left(\frac{n\,\omega^2}{g}
    -k^2\,z\right)F=0.
\end{equation}
The EOS of a strange star crust can be
well approximated by the polytropic
EOS with the index $n=3$. Thus,
Eq.\ (\ref{F_polytrop}) with $n=3$ will be used below for
an analytical consideration of the oscillation problem.

The frequencies $\omega$ refer to the local crust reference frame
(see Eq.~(\ref{metric})). The frequencies $\widetilde\omega$ of
oscillations detected by a distant observer are
\begin{equation}
       \widetilde\omega=\omega \, \sqrt{1-R_{\rm G}/R} .
\end{equation}
%

\subsection{Radial oscillations}
\label{SubSecRadOsc}

For radial oscillations ($l=0$) Eq.\
(\ref{F_polytrop}) is simplified,
\begin{equation}
    z\,{\frac{{\rm d} {^2F}}{{\rm d} {z^2}}}+n{\frac{{\rm d} {F}}{{\rm d} {z}}}
    +\frac{n\,\omega^2}{g}\,F=0.
\end{equation}
The solution bounded at the surface ($z=0$) is
\begin{equation}
\label{RadModes}
    F=A\,J_{n-1}\left(
    2\,\omega\,\sqrt{n\,z/g}\right) / z^{(n-1)/2},
\end{equation}
where $A$ is a constant, and $J_n(x)$  is a Bessel function of the
first kind \citep[see, e.g.,][]{Abramowitz}. Then the
eigenfrequencies are given by the equation
\begin{equation}
\label{EqRadFeq}
    J_n\left(2\,\omega\,\sqrt{n\,h/g}\right)=0.
\end{equation}
Formally, the solution contains the eigenfrequency $\omega=0$ which
corresponds to $F(z)\equiv {\rm const}$ (i.e., to a meaningless
vanishing oscillation velocity). In addition, the equation gives an
infinite number of eigenfrequencies,
\begin{equation}
\label{RadFreq}
   \omega_i=\frac{j_{n,\,i}}{2}\,\sqrt{\frac{g}{n\,h}},
\end{equation}
where $j_{n,\,i}$ is an $i$th zero of $J_n(x)$. For a given $n$ and
a given maximum density in the crust, $\rhomax$, the crustal depth
$h\propto g^{-1}$, and $\omega$ is proportional to $g$, in agreement
with the general relation (\ref{scaling}). Note, that a mode labeled
by $i$ has $i$ nodes of $F(z)$. At $n=3$ we have $j_{3,\,i}=
6.38,\,9.76,\,13.0,\,16.2,\,19.4$ for $i$ from 1 to 5. For $i\geq 2$
the difference of neighboring values is
$j_{3,\,i+1}-j_{3,\,i}\approx 3.2$ (and it is slightly larger
$\approx 3.38$ for $i$=1). Accordingly, the eigenfrequencies are
approximately equidistant. This feature can be used to distinguish
such oscillations from other ones (see Section
\ref{SubSecStarParams}).

\subsection{Nonradial oscillations}
\label{SubSecNonRadOsc}

The problem of non-radial oscillations of a thin polytropic layer
was studied by \citet{Lamb,LambBook}. Following those studies let us
introduce $f(z)=\exp(kz)\,F(z)$ and $a=n\,(1-\omega^2/gk)/2$ in
Eq.~(\ref{F_polytrop}). Then Eq.~(\ref{F_polytrop}) reduces to
\begin{equation}
    z\,\frac{\partial^2 f}{\partial z^2}
    +(n-2\,k\,z)\frac{\partial f}{\partial z}
    -a\,f=0.
\end{equation}
The solution to this equation bounded at $z\rightarrow 0$ is
$f(z)=A\,M(a,n,2kz)$, where $A$ is a constant and $M(a,b,x)$ is the
Kummer function \citep[see, e.g.,][]{Abramowitz}. The boundary
condition $F^\prime(h)=0$ leads to the equation
\begin{equation}
\label{Polytrop_bound}
 \frac {2\,a}{n}\,M(a+1,n+1,2kh)=M(a,n,2kh).
\end{equation}
For a not very large multipolarity $l\ll R/h\sim 50$, the value of
$kh$ is small,  and the long-wavelength approximation applies which
simplifies problem. The eigenfrequency of the fundamental mode can
be determined using asymptotic series expansions, which give
\begin{equation}
\label{FundFreq}
    \omega^2_{\rm f}=k^2\,gh/(n+1)=l\,(l+1)\,g\,h/\left[R^2\,(n+1)\right].
\end{equation}
The respective eigenfunction is $F_{\rm
f}=A\,(1+k^2\,(z^2/2-hz)/(n+1))$ \citep{Lamb,LambBook}. In this case
matter elements move nearly horizontally; the ratio of the radial to
the horizontal velocity components can be estimated as $kh/(n+1)\ll
1$.

The p modes can be studied numerically. However, they have
relatively high frequencies, so that $k^2 \ll \omega^2/c_s^2$ in
Eq.\ (\ref{F}). Neglecting $k^2$, we come to the same equation as
for radial oscillations. Thus, higher modes of non-radial
oscillations have almost the same frequencies (\ref{RadFreq}) and
radial parts (\ref{RadModes}) of the velocity potential $F(z)$, as
radial oscillations (see Section \ref{SubSecRadOsc}). These results
can also be obtained using the asymptotes of the Kummer function. In
these cases the velocity is mostly radial; the ratio of the radial
to the horizontal velocity components can be estimated from
Eq.~(\ref{F_polytrop}) as $1/kh$.

\subsection{Oscillation damping}

We define the oscillation damping time (in the local
stellar reference frame) as
\begin{equation}
\label{tau}
    \tau=E / |{\rm d}E/{\rm d}t|,
\end{equation}
where the oscillation energy is
\begin{equation}
\label{E}
    E=
     \frac {R^2}2 \int_0^{h} \rho \left[\left(F\,^\prime\right)^2
         +k^2\,F^2\right]\,{\rm d}z
\end{equation}
and
\begin{equation}
   \frac{{\rm d}E}{{\rm d}t} =\frac{{\rm d}E_{\rm el}}{{\rm d}t}
    +\frac{{\rm d}E_{\rm grav}}{{\rm d}t}
    +\frac{{\rm d}E_{\rm visc}}{{\rm d}t}.
\end{equation}
Here, ${\rm d}E_{\rm el}/{\rm d}t$, ${\rm d}E_{\rm grav}/{\rm d}t$
and ${\rm d}E_{\rm visc}/{\rm d}t$ are oscillation energy loss rates
owing to the emission of electromagnetic and gravitation waves and
owing to the viscous dissipation.

To estimate the gravitational radiation rate (for $l\ge 2$) we
employ the multipole expansion formula \citep{Balbinski}
\begin{equation}
\label{grav_zatuh}
 \frac{{\rm d}E_{\rm grav}}{{\rm d}t}=2\pi\,
 \frac{l\,(l+1)\,(l+2)}{(l-1)\left[(2l+1)!!\right]^2}\,
  \frac{GR}{\omega}\,
 \left(\frac{\omega\,R}{c}\right)^{2l+1}\,
 \,I_{\rm grav}^2,
\end{equation}
where
\begin{equation}
\label{Igrav}
 I_{\rm grav}=
      \int_0^h \rho \left(F^\prime(z)+(l+1)F(z)/R\right)\,{\rm d} z.
\end{equation}

The electromagnetic damping rate (for $l\ge 1$) calculated in the
model of a frozen-in dipolar magnetic field disturbed by
oscillations, with the vacuum boundary conditions at the stellar
surface \citep{McDermott,Tsygan}, is given by
\begin{eqnarray}
   l=1\quad \frac{{\rm d}E_{\rm el}}{{\rm d}t}&=&
        \frac{c}{720\,\pi} B^2\, R^2\, \left(\frac{\omega\,R}{c}\right)^6\,
        \left(\frac{2\,F(0)/R+F^{\prime}(0)}{R\omega}\right)^2;
   \\
   l\ge 2\quad  \frac{{\rm d}E_{\rm el}}{{\rm d}t}&=&\frac{c}{32\,\pi}\,B^2\,R^2\,
         \left(\frac{\omega\,R}{c}\right)^{2l}\,\frac{ \,l\,(l-1)}{(2\,l+1)\,
     (2\,l-1)\,\left[(2l-3)!!\right]^2}
\nonumber \\
      &\times&
      \left\{\frac{2\,(l+1)\,F(0)/R-F^{\prime}(0)}{R\,\omega}\right\}^2,
   \label{elmag_zatuh}
\end{eqnarray}
where $B$ is the magnetic field strength at the magnetic pole. The
presence of the magnetosphere can change the emission power
\citep*{Timokhin}, but we neglect this effect.

The viscous damping of oscillations was studied by \cite{Chug2005}
who showed that
\begin{equation}
\label{ViscDump}
    \frac {{\rm d}E_{\rm visc}}{{\rm d}t}
    =-\frac 14 \int_0^{R} r^2\, \eta
     \left(I_1-\frac 43 I_2\right)\,{\rm d} r,
\end{equation}
where
\begin{eqnarray*}
I_1&=&4\,\left\{\left(F^{\prime\prime}\right)^2
    +2\,\frac{1+l(l+1)}{r^2}\,\left(F\,^\prime\right)^2
    -6\,\frac{l(l+1)}{r^3}\,F\,^\prime F
    +l(l+1)\,\frac{1+l(l+1)}{r^4}\,F^2\right\} ,
\\
I_2&=&\left(F^{\prime\prime}+\frac{2\,F\,^\prime}{r}
    -\frac{l(l+1)}{r^2}\,F\right)^2,
\end{eqnarray*}
and $\eta$ is the shear viscosity.
Note, that the viscous damping can be
additionally enhanced by thin viscous layers
near weak first-order phase transitions
associated with transformation of nuclides
in dense matter. The viscosity in these layers can be
diffusive or turbulent; we neglect this additional dissipation in our
calculations.

Taking the equilibrium pressure $P$ as the integration variable in
Eq.~(\ref{E}) we obtain
\begin{equation}
  E=\frac{g\,R^2}{2}\int_0^{P_{\rm
  max}}\left(\rho^2\,{F^\prime_P}^2+\zeta^2\,F^2\right)\,{\rm d}P,
\end{equation}
where $P_{\rm max}$ is the pressure at the crust bottom, and
$F^\prime_P={\rm d}F/{\rm d}P$. Note, that for given EOS,
$\rhomax$, $\zeta$, and the root-mean-square radial displacement at
the surface, the energy scales as $g\,R^2$.

As will be shown in Section \ref{SecNumRes}, the viscous damping
dominates at $l=0$ and $l\gtrsim 30$. Dipole modes
decay usually owing to electromagnetic radiation.
Depending on the magnetic field strength, modes with
$2\le l\lesssim 30$ damp mainly under the action of either
electromagnetic or gravitational radiation.

Let us consider the oscillation damping for a polytropic EOS
for which the equilibrium density is
$\rho=\rhomax\,\left(z/h\right)^n$.
In this case, the damping rate via gravitational radiation is
evaluated analytically and the oscillation energy
(for a given rms amplitude of radial surface displacements)
is independent of $l$. The results are outlined below.

\subsection{The damping of f modes}

The oscillation damping of a fundamental mode comes mainly from
horizontal motions of the matter. Accordingly, it is sufficient to
take $F_{\rm f}=A\,(1+k^2\,(z^2/2-h\,z)/(n+1))\approx A$. Then
\begin{equation}
    I_{\rm grav}=A\,\rhomax\,\frac{h}{R\,(n+1)}
\end{equation}
and
\begin{equation}
   E=\frac{l(l+1)}{2}\,\rhomax\,A^2\,h.
\end{equation}
Note that for a given root-mean-square radial displacement $a$
at the surface, the
normalization constant is $A\propto a\,\omega /k^2\propto a/k$
(as follows from Eq.~(\ref{FundFreq})). Accordingly, the
energy is independent of $l$.

\subsection{The damping of p modes} \label{SubSecPmodeDump}

P modes (with at least one radial node) of not very large
multipolarity $l\ll R/h$ are accompanied by nearly radial velocities
of the matter. Their velocity potential $F$ and oscillation
frequencies are almost the same as for radial oscillations. Thus,
their energy can be estimated as
\begin{equation}
  E\approx \frac{R^2}{2}\int_0^{h}\rho\,{F^\prime}^2\,{\rm d}z.
\end{equation}
For the same root-mean-square radial displacement on the surface
the energy is independent of $l$. One has
\begin{equation}
\label{Igrav_higher}
 I_{\rm grav}\approx
      \int_0^h \rho F^\prime(z)\,{\rm d} z
      =-A^2\,h^{(1-n)/2}\,\rho_{\rm
      max}\,J_{n+1}\left(2\,\omega\sqrt{ h\,n / g}\,\right);
\end{equation}
the last equality is obtained using Eq.\ (\ref{RadModes})
for the function $F$.

\section{Numerical results}
\label{SecNumRes}

All numerical results are presented for a strange star of the
gravitational mass $M=1.4\,M_\odot$, the circumferential radius
$R=10$~km, and the crust depth $h=250$~m. The results can easily be
rescaled to other strange star models (with the same EOS and the
same maximum density of the normal matter) using the scaling
relation (\ref{scaling}).

Eigenfrequencies have been determined from Eq.\ (\ref{F})
(with the boundary conditions
$F^\prime(h)=0$ and $F$ bounded at $z \rightarrow 0$)
by the Runge-Kutta method.

\begin{figure}
    \begin{center}
        \leavevmode
        \epsfxsize=120mm \epsfbox[33 34 529 383]{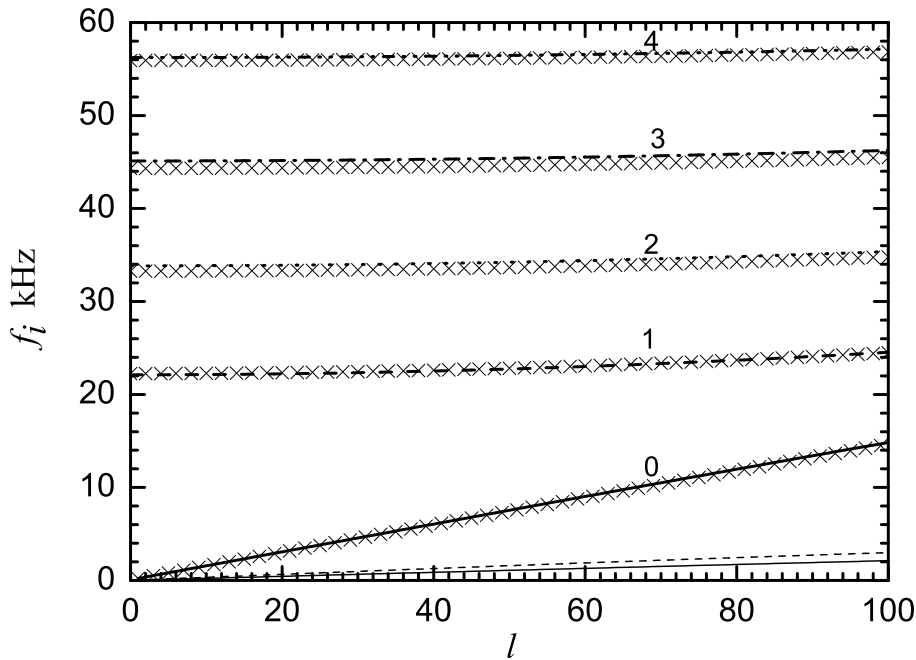}
    \end{center}
    \caption{Frequencies of oscillations localized in the crust
         of a strange star ($M=1.4\,M_\odot$, $R=10$~km and $h=250$~m)
     as detected by a distant
         observer. The numbers next to
         the curves indicate the number of radial nodes of the function $F$.
         Crosses are plotted for the EOS of the accreted crust
         \citep{HZ}; lines are for the polytropic EOS
     with $n$=3.
         The two oscillation branches of the accreted crust
     (shown by thinner lines)
         with lowest frequencies
         are density discontinuous g modes.
    }
    \label{Fig_Freq}
\end{figure}

Figure \ref{Fig_Freq} shows oscillation eigenfrequencies calculated
for a distant observer as a function of multipolarity $l$. Crosses
are plotted for the EOS of the accreted matter \citep{HZ} with the
accurate treatment of phase transitions \citep[see,
e.g.,][]{Chug06}. Lines are for the polytropic EOS ($n=3$), which
describes the normal matter composed of $^{56}$Fe nuclei and
ultrarelativistic electrons. The density at the crust bottom is
$\rhomax=7.65\times 10^{10}$~g~cm$^{-3}$ for the accreted crust and
$3.9\times 10^{10}$~g~cm$^{-3}$ for the polytropic one. Note, that
the polytropic EOS accurately describes most of eigenfrequencies,
but oscillations of the accreted crust have two specific additional
branches (shown by thinner lines in Figure \ref{Fig_Freq}) known as
density discontinuity g modes \citep[see, e.g.,][]{McDermott1990}.
These oscillations are caused by buoyancy forces associated with
phase transitions; they are present in neutron stars and in strange
stars with crust. For a vanishing shear modulus of the matter in the
vicinity of a phase transition, the frequencies of such modes can be
estimated as \citep{McDermott1990}
\begin{equation}
    f\approx 1.8
     \left\{ l(l+1)\, \frac{1-\exp(-n)}{n}
            \left(1-\frac{R_{\rm  G}}{R}\right)
     \right\}^{1/2}\,
     \left(\frac{10\,{\rm km}}{R}\right)^{3/2}\,
     \left(
            \frac{M}{M_\odot}\,
            \frac{\Delta\rho}{\rho_{-}}\,
            \frac{z_{\rm  ph}}{R}
     \right)^{1/2}\,{\rm kHz},
\end{equation}
where $z_{\rm ph}$ is the depth of the phase transition, $\Delta
\rho$ is the density jump, and $\rho_{-}$ is the density just after
the jump. The parameters of such modes are sensitive to the model of
accreted matter, to the shear modulus of this matter, etc. Note that
the distinctness of phase transitions in the accreted crust is still
not clear
--- they could be smoothed (which would affect
the respective g mode oscillation frequencies). Therefore, we will
not include density discontinuity g modes in our analysis and will
study other oscillation branches in the polytropic approximation.

\begin{figure}
    \begin{center}
        \leavevmode
        \epsfxsize=120mm \epsfbox[33 34 529 383]{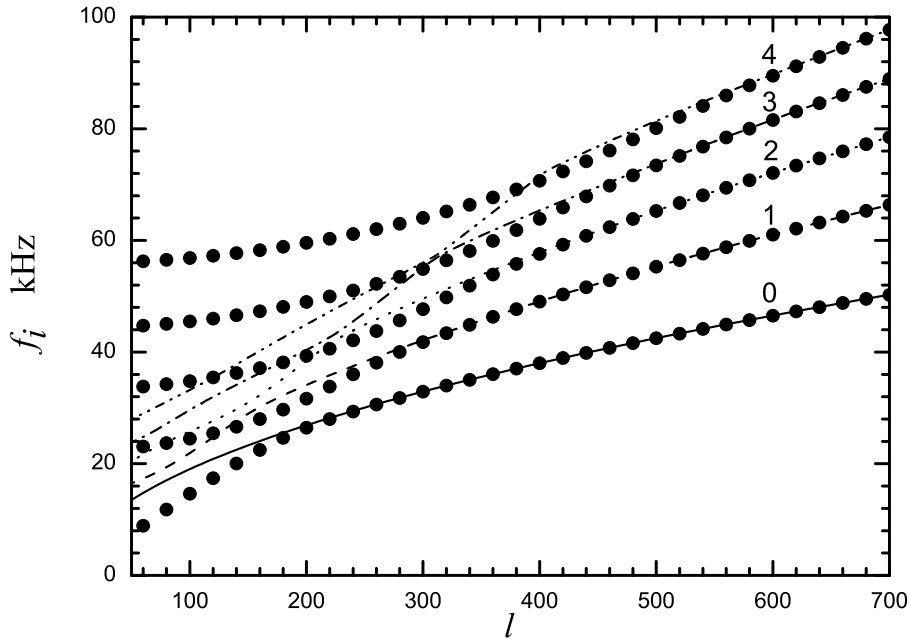}
    \end{center}
    \caption{Frequencies of oscillations
         (as detected by a distant observer)
     localized in crusts
         of strange stars (filled dots) and neutron stars (lines) of
         the same mass $M=1.4\,M_\odot$ and radius $R=10$~km.
         For the strange star, the crust depth is chosen to be $h=250$~m.
         The density discontinuous g modes
         are not shown. Numbers near curves show the
     number of radial nodes of the function $F$.
    }
    \label{Fig_SSvsNS}
\end{figure}

Figure \ref{Fig_SSvsNS} shows the frequencies of oscillations of a
strange star and a neutron star of the same mass $M=1.4\,M_\odot$
and radius $R=10$~km.  Both stars are assumed to have an accreted
crust. For the strange star, the crust depth is chosen to be
$h=250$~m. Details of calculations of neutron star
oscillations are described by
\cite{Chug06}. If $l \gtrsim 300$, oscillations are localized in a
thin surface layer of the depth $\lesssim h$; then both the strange
star and the neutron star have the same oscillation spectrum. For
lower $l$ the oscillations can penetrate into the layers of the
depth $\gtrsim h$, and the oscillation spectra of the strange star
and the neutron star are seen to become different.

Let us focus on oscillations of the crust of strange stars. For not
very large $l\lesssim 150$ the frequencies of fundamental modes
depend linearly on $l$, but for p modes (with at least one radial
node) the frequencies are approximately constant and equidistant.
This feature is in good agreement with analytical results of
Sections \ref{SubSecRadOsc} and \ref{SubSecNonRadOsc}. For higher
$l$ the $l$-dependence of the frequencies becomes more complicated
because of the violation of the long-wave approximation (which
requires $kh\ll 1$). Finally, for $l\gtrsim 300$, the oscillations
are localized in the outer layers of the crust, and their
frequencies become the same as for neutron stars.

Oscillations of strange stars with crust have several specific
features which can be used to distinguish them from oscillations of
neutron stars.
\begin{itemize}
\item The frequencies of fundamental modes have linear dependence on
$l$.
\item The frequencies of p modes (with at least one radial
node) are approximately equidistant and have very weak dependence
on $l$.
%
%
\end{itemize}

%
\begin{figure}
    \begin{center}
        \leavevmode
        \epsfxsize=120mm \epsfbox[43 35 547 401]{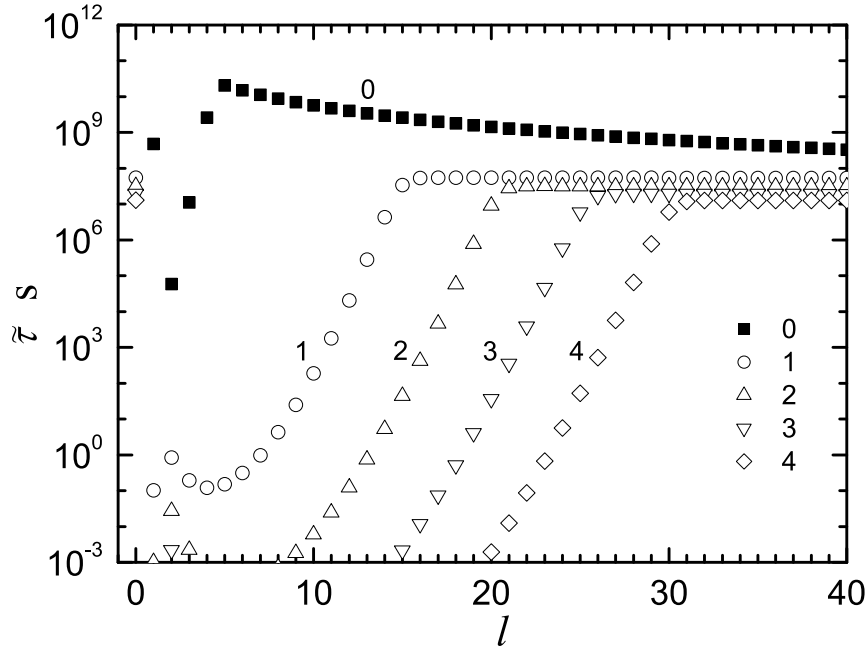}
    \end{center}
    \caption{Damping times of oscillations (for a distant
         observer) localized in the crust
         of a strange star ($M=1.4\,M_\odot$, $R=10$~km, and $h=250$~m).
         The crust temperature is $T=10^7$~K and
         the magnetic field at the magnetic poles
         is $B=10^{12}$~G. Numbers 
         indicate the number of radial nodes of the function $F$.
    }
    \label{Fig_Dump}
\end{figure}

 Figure \ref{Fig_Dump} shows the damping time of
oscillations, rescaled for a distant observer
($\widetilde{\tau}=\tau\,(1-R_{\rm G}/R)^{-1/2}$). The crust
temperature (important for the viscous damping) is assumed to be
$T=10^7$~K and the magnetic field strength at the magnetic poles is
$10^{12}$~G. For higher $T$ the viscous damping time becomes larger.

The damping time is seen to vary by many orders of magnitude, from
$\widetilde{\tau}\approx 10^3$~years for the fundamental mode with
$l=5$ to $\widetilde{\tau}\lesssim 10^{-3}$~s for the mode with at
least 2 radial nodes of $F(z)$ and $l\sim 10$. This huge difference
is produced by several damping mechanisms.
In particular, radial oscillations ($l=0$) decay exclusively through
the viscous dissipation (generating neither gravitational
nor electromagnetic radiation). They damp slowly, with
$\widetilde{\tau}\sim 1$~year.
The modes with $l=1$
undergo powerful electromagnetic damping which
greatly decreases the damping time (Figure \ref{Fig_Dump}).

Fundamental modes have low frequencies. Accordingly, they mainly
undergo the viscous damping and do not damp efficiently via the
emission of electromagnetic and gravitation waves. Their damping
times are relatively large, up to $10^3$~years. The exclusion is
provided by the modes with $l=1,\,2,\, 3$ and 4, which decay quicker
via the emission of electromagnetic waves.

For p modes (with at least one radial node) the picture is
different. The modes with $1\le l\lesssim 30$ decay primarily though
the electromagnetic channel.
Because neither oscillation frequency nor $F(z)$ depend on $l$
(see Section
\ref{SubSecNonRadOsc}),
the $l$-dependence of the dissipation time
is fully determined by the electromagnetic
energy losses (see Eq.\ (\ref{elmag_zatuh})). For
higher multipolarity $l\gtrsim 30$, the electromagnetic and
gravitational emissions are strongly suppressed by large $l$ (see
Eqs.~(\ref{grav_zatuh}) and (\ref{elmag_zatuh})) and the viscous
damping prevails. As a result, the damping times become
approximately constant due to a weak $l$-dependence of oscillation
frequencies and radial parts $F(z)$ of the velocity potential. Note
that for lower $B\sim 10^9$~G, the main damping mechanism of modes
with $2\le l\lesssim 20$ consists in the emission of gravitational
waves, while the electromagnetic channel is important only for
dipole modes.

To conclude, the fundamental modes and all radial oscillations have
damping times exceeding 1 year and have better chances to be
detected. The modes with $l\gtrsim 30$ have also relatively large
damping times $\gtrsim 1$ year, but it could be difficult to excite
and detect them owing to large multipolarity.

\subsection{Inferring strange star parameters}
\label{SubSecStarParams}

Let us assume that some oscillation frequencies are detected and
corresponding modes are identified. The frequencies of pressure
modes (in the local stellar reference frame)
 are given by Eq.\ (\ref{RadFreq}). Then the measured
frequencies $\widetilde{\omega}_i$ would give us the value of
\begin{equation}
\label{B}
    \alpha=\frac{g}{h}\,\left(1-R_{\rm G}/R\right)
    =3\,\left(\frac{2\,\widetilde{\omega}_i}{j_{3,\,i}}\right)^2.
\end{equation}
The frequencies of fundamental models are given by Eq.\
(\ref{FundFreq}). If $\widetilde{\omega}_f$
were measured, it would provide
us with the value of
\begin{equation}
\label{C}
    \beta=\frac{g\,h}{R^2 }\,\left(1-R_{\rm G}/R\right)=4\,\frac{\widetilde{\omega}_{\rm f}^2}{l(l+1)}.
\end{equation}
Formally, we need to detect only one mode of each type to infer the
values of $\alpha$ or $\beta$, but detecting several modes would
give more confidence to the results.

Having $\alpha$ and $\beta$, we could obtain the value of
$ 
    \gamma=\sqrt{\alpha\,\beta}=g\,\left(1-R_{\rm G}/R\right)/R
$ 
 without any additional assumption on the value of
$\rhomax$. Note that $g\,\sqrt{1-R_{\rm G}/R}/R\propto \bar{\rho}$,
the mean density of the star.

Let us assume
that we can deduce the value of $\rhomax$
form the theory or from interpretation of some observations.
Then the crust depth $h$ could be calculated for a given EOS. For the
polytropic EOS it can be estimated as
\begin{equation}
\label{h}
    h\approx 8.31\times 10^4\, (\rhomax_{11})^{1/3}/g_{14}\mbox{~cm},
\end{equation}
where $g_{14}=g/10^{14}$~cm\,s$^{-2}$ and
$\rhomax_{11}=\rhomax/10^{11}$~g\,cm$^{-3}$. In that case we would
obtain two equations (for $\alpha$ and $\beta$) for two important
unknown stellar parameters, the mass $M$ and the radius $R$, and
could determine these parameters. If we do not know $\rhomax$, we
would be unable to determine $M$ and $R$, but could constrain them
taking into account that $\rhomax$ cannot exceed the neutron drip
density $\rho_{\rm d}\approx 6\times 10^{11}$~g\,cm$^{-3}$ (for the
accreted envelope). Substituting $\rhomax=\rho_{\rm d}$ into
(\ref{h}) and $h$ into (\ref{C}) we would obtain the upper limit on
$R$. With this upper limit, we could easily derive the upper limit
on $M$ from the value of $\gamma$.

If, on the other hand, the mass $M$ is known from independent
observations (e.g., the star enters to a binary system) then the
equations for $\alpha$ and $\beta$ would allow one to determine $R$
and $h$.

Note, that we do not include density discontinuous g modes into our
analysis. The frequencies of these modes depend on many factors
(such as the densities of phase transitions, associated density
jumps, etc.). The uncertainties of such factors would complicate the
theoretical interpretation of measured g mode frequencies. However,
if observed, g modes could provide some useful additional
information. For instance, the number of such modes for a fixed $l>
0$ is equal to the number of phase transitions $N$ (this statement
is strict only in the absence of other sources of buoyancy). This
number could bound the maximum crust density $\rhomax$ in the region
between theoretically predicted $N$-th and $N+1$-th phase
transitions and impose then better bounds on the inferred values of
$M$ and $R$.

\section{Conclusions}

We have studied pressure oscillations of
strange star crusts.

Our main conclusions are as follows.

(1) The oscillations are almost insensitive to the various
modifications of the EOS in the normal crust. The polytropic EOS
provides approximately the same eigenfrequencies as the EOS of the
accreted crust, except for density discontinuous g modes, which are
absent for the polytropic EOS (see Section \ref{SecNumRes} and
Figure \ref{Fig_Freq}).

(2) The oscillation problem for acoustic modes is self-similar (in
the plane-parallel approximation). Once the problem is solved for
one stellar model, it can easily be rescaled to strange star models
with any mass and radius (but the same EOS and the maximum crust
density; see Section \ref{SecFormal}).

(3) For a thin polytropic crust,
the oscillation problem is solved analytically (Sections
\ref{SubSecNonRadOsc} and \ref{SubSecRadOsc}).

(4) The oscillation spectrum of a strange star crust is {\it
specific}. The frequencies of fundamental modes depend linearly on
$l$; the frequencies of p modes are almost independent of $l$. These
features are {\it unmistakable seismic signatures of strange stars
with crust} (Section~\ref{SecNumRes}).

(5) A detection and identification of one fundamental mode and one p
mode would enable one, in principle, to infer the mass and radius of
a strange star (if the maximum crust density is known) or at least
to obtain corresponding upper limits (Section
\ref{SubSecStarParams}).

Therefore, oscillation modes of strange stars with crust are potentially good
tools to distinguish these strange stars from neutron stars and to
determine their masses and radii. The oscillation frequencies could be
detected by radio-astronomical methods very precisely.

A search for these oscillation modes could be useful. Some of
them do not damp quickly and can survive
for a long time (Section \ref{SecNumRes}). Pressure
modes are robust because they are relatively independent of the
thermal state of the crust, and they should not be strongly affected
by the crustal magnetic field.

\section*{Acknowledgments}

I am grateful to D.G. Yakovlev for discussions and
also to V.A. Urpin and the anonymous
referee for useful critical remarks. This work was supported by a grant of the
''Dynasty'' Foundation and the International Center for Fundamental
Physics in Moscow, by the Russian Foundation for Basic Research
(project no.\ 05-02-16245) and
by the Federal Agency for Science and Innovations
(grant NSh 9879.2006.2).

                    

\label{lastpage}
\end{document}